\documentclass[twocolumn,showpacs,preprintnumbers,superscriptaddress]{revtex4}
\usepackage{graphicx}
\usepackage{dcolumn}
\usepackage{bm}
\usepackage{color}
\usepackage{amsmath}
\usepackage{amssymb}
\usepackage{amsfonts}

\begin{document}

\title{Simulation and detection of photonic Chern insulators in one-dimensional circuit QED lattice}

\author{Feng Mei}
\email{tianfengmei@gmail.com}
\affiliation{National Laboratory of
Solid State Microstructures, Department of Physics, Nanjing
University, Nanjing, China}
\affiliation{Centre for Quantum
Technologies, National University of Singapore, 3 Science Drive 2,
Singapore 117543, Singapore}

\author{Jia-Bin You}
\affiliation{Centre for Quantum
Technologies, National University of Singapore, 3 Science Drive 2,
Singapore 117543, Singapore}

\author{Wei Nie}
\affiliation{Centre for Quantum
Technologies, National University of Singapore, 3 Science Drive 2,
Singapore 117543, Singapore}

\author {Rosario Fazio}
\email{rosario.fazio@sns.it}
\affiliation{Centre for Quantum Technologies, National University of Singapore,
3 Science Drive 2, Singapore 117543, Singapore}
\affiliation{NEST, Scuola Normale Superiore and Istituto Nanoscienze-CNR, I-56126 Pisa, Italy}

\author{Shi-Liang Zhu}
\email{slzhu@nju.edu.cn}
\affiliation{National Laboratory of Solid
State Microstructures, Department of Physics, Nanjing University,
Nanjing, China}
\affiliation{Synergetic Innovation Center of Quantum Information and Quantum Physics,
University of Science and Technology of China, Hefei 230026, China}

\author{L. C. Kwek}
\email{cqtklc@nus.edu.sg}
\affiliation{Centre for Quantum Technologies, National University of Singapore,
3 Science Drive 2, Singapore 117543, Singapore}
\affiliation{Institute of Advanced Studies, Nanyang Technological University, 60 Nanyang View, Singapore 639673}
\affiliation{National Institute of Education, Nanyang Technological University, 1 Nanyang Walk, Singapore 637616}

\date{\today}

\begin{abstract}
We introduce a conceptually simple and experimentally feasible method to realize and detect photonic
topological Chern insulators with one-dimensional circiut quantum
electrodynamics lattice. By periodically modulating the couplings
in this lattice, we show that this one-dimensional model can be
mapped into a two-dimensional Chern insulator model. In addition
to allowing the study of photonic Chern insulators, this approach
also provides a natural platform to realize experimentally
Laughlin's pumping argument. Remarkably, based on scattering theory of
topological insulators and input-output formalism, we find that
both photonic edge state and topological invariant can be unambiguously probed with a simple
dissipative few-resonator circuit-QED network.
\end{abstract}


\maketitle

 \emph{Introduction.} The recent rapid experimental developments in circuit quantum electrodynamics (circuit QED)
 have turned this system into one of leading platforms for studying quantum optics and quantum computation \cite{Devoret2013}.
 This system possesses high coherence superconducting qubits \cite{Martinis2014} and well controllable coupling parameters \cite{tworesthe,tworesexp}.
 Putting qubits and microwave resonators into a lattice further allows this system to be used for solid state quantum simulation. Quantum state can be
 easily manipulated and detected at single-site level in such lattice. Combined with on-site nonlinearity or photon blockade, circiut QED lattices have
 been widely explored as quantum simulators in the past years for investigating photon- or polariton-based many-body physics \cite{CircuitQS,PolaritonQS,Rosario}. Experimentally, mimicking quantum spin models with qubit arrays recently has been successfully demonstrated \cite{DQS}.

 Topological photonics nowadays has become a very active area of research \cite{topopho}. Photonic topological insulator was firstly predicted in
  a two-dimensional photonic crystal \cite{Haldane} and subsequently has been extensively studied \cite{Marin1,Marin2,Mikael,Alexander,Chan}.
  Based on engineering artifical magnetic field and spin-orbit coupling, photonic quantum integer and spin Hall states have also been studied in two dimensional coupled resonators \cite{Hafezi2013,Hur,Hafezi2011,Chong,Girvin,Carusotto2011,Fan2012,Ripoll}
  and linear circuit lattice \cite{Jiang,Simon}. Experimentally, the propagation of photonic chiral edge states has been
  observed \cite{Marin2,Mikael,Chan,Hafezi2013,Kraus}. It would be highly desirable, in these experiments,
  to measure the photonic topological invariants. Due to the different statistics of the carrier, photonic topological invariants cannot be measured
  using the same methods devised for electronic systems. Recently, how to detect photonic topological invariants has been studied based on probing
  the Berry curvature \cite{Carustotto2014} and the dynamics of edge states
  \cite{Hafezi2014,Hafezi20151}. However, although both circuit QED and topological photonics have been rapidly developed,
  the connection of them is less studied.

\begin{figure}[h]
\includegraphics[width=8.5cm,height=4cm]{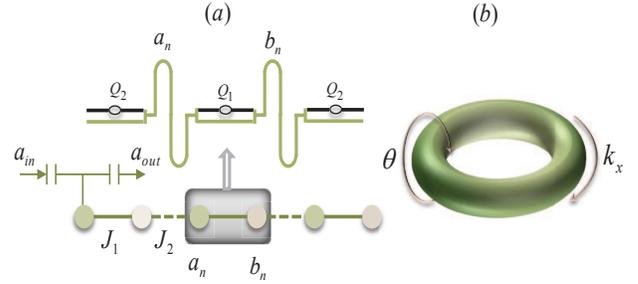}
\caption{(a) Setup for the one-dimensional circuit QED lattice,
where the cavity input-output process has been explored to probe the edge state and topological
invariant. (b) An artificial torus is constructed by combining the
parameter $\theta$ with the quasimomentum $k_x$ of the photonic
lattice.}
\end{figure}

 In this paper, we propose a conceptually simple method to simulate a two-dimensional photonic topological Chern insulator
 with a one-dimensional transmission line resonator photonic lattice. By identifying a periodic parameter introduced in the system as the quasi-momentum in a second artificial dimension, we show that this one-dimensional photonic lattice can be mapped onto a two-dimensional lattice exhibiting
 the Chern insulator state.
 Compared to previous methods of engineering photonic topological insulators, our method employs a one-dimensional photon lattice, and it does not require mimicking gauge fields and spin-orbit coupling, which opens a new and simple route to study and probe high dimensional photonic topological states. It is also interesting to note that such one-dimensional system can provide a natural platform to realize Laughlin's pumping argument\cite{Laughlin} and to test scattering theory of topological insulators \cite{scatp}. Based on such feature, we generalize such scattering theory to take into account dissipation and demonstrate that both photonic edge states and topological invariant still can be clearly probed from the final steady state. We believe that our results could simulate more studies on the dissipation effect in topological states. It is also noteworthy to stress that the topological features emerge even in a dissipative few-resonator circuit-QED network. Topological invariant defined on a two-dimensional topological lattice can be unambiguously detected with a small-size one-dimensional simulator.

\emph{One-dimensional circuit QED lattice.}  We start with a transmission line resonator lattice, as shown in Fig. 1(a).
The photon hopping between nearest neighbour resonators is mediated through the coupling capacitors and the connected flux qubits.
Note that this coupling method has recently been demonstrated in the two-resonator circuit QED experiments \cite{tworesthe,tworesexp}.
In both cases, the coupling has been designed to make sure that the resonator lattice has the alternating hopping configuration.
Each unit cell of this lattice has two resonators, labelled by $a$ and $b$. The capacitively coupled resonator lattice is described by the Hamiltonian
\begin{equation}
H_0=\sum_{n}J_1a^{+}_nb_n+J_2a^{+}_{n}b_{n-1}+h.c.,
\end{equation}
where $J_1$ and $J_2$ are the intra- and inter-cell hopping rates.

For the qubit-assisted hopping, we assume that the two resonators within the same unit cell are both coupled to the flux qubit $Q_1$,
while the two resonators belonging to the two nearest-neighbour unit cells are both coupled with the flux qubit $Q_2$. The purpose of this
coupling is to provide an alternating parametric modulation on the hopping rates and the on-site energies.  In the dispersive regime, when
all the qubits are in the ground state, the coupling between the resonator and the qubit can be removed, leading to an effective transmission
resonator lattice with photon hopping assisted by the connected qubits. Combined with the previous capacitively coupled resonator lattice, the total Hamiltonian of this cicuit-QED lattice
(in a rotating frame with respect to the external driving frequency $\omega_d$ and also in the interaction picture with respect to the
qubit energy $\omega_{1,2}$) takes the form
\begin{equation}
\begin{aligned}
H=\sum_n(J_1-\frac{g_1g_2}{\Delta})a^{+}_nb_n+(J_2+\frac{g_1g_2}{\Delta})a^{+}_{n}b_{n-1}+h.c\\
+\frac{g^2_2-g^2_1}{\Delta}(a^{+}_{n}a_{n}-b^{+}_{n}b_{n})+\Delta_c(a^{+}_{n}a_{n}+b^{+}_{n}b_{n}),
\end{aligned}
\end{equation}
where $g_1$ and $g_2$ describe the coupling strengths between the qubit $Q_1 (Q_2)$ and the resonators $a_n$ and $b_n$,
($b_n$ and $a_{n+1}$ ), $\Delta=\omega_1-\omega_d= \omega_d-\omega_2$ is the detuning of the qubit energies, and
$\Delta_c=\omega_c-\omega_d$ is the detuning of the resonator frequency.
The qubit-assisted hopping and on-site modulation terms are introduced in order to map into the second dimension for the
construction of the photonic Chern insulator.

\emph{Two dimensional lattice mapping.}  To simulate the two-dimensional Chern insulator Hamiltonian \cite{TIrev},
we write the qubit-resonator coupling strengths in the above lattice Hamiltonian in a parameter space as
\begin{equation}
g_1=g_0\sin(\theta/2), g_2=g_0\cos(\theta/2),
\end{equation}
where the mixing angle $\theta=2\arctan(g_1/g_2)$ and $g_0=\sqrt{g^2_1+g^2_2}$. The parameter $\theta$ is determined by the ratio between the coupling strength $g_1$ and $g_2$. Note that the coupling strengths between the flux qubit and the resonators can be individually controlled through using superconducting quantum interferences (SQUIDs) devices and changing the external magnetic fluxes applied on the SQUIDs loops \cite{Solanosm}. Then $\theta$ can be engineered from 0 to $2\pi$ for subsequent two dimensional mapping. Moreover, the topological feature demonstrated below in this model endows this system with topological protection, which allows our methods to be robust to practical deformations in the parameters engineering. By substituting the above equation into the total lattice Hamiltonian and further writing it in momentum space, one can get $H=\sum_kC^{+}_kh(k)C_k$, where
$C_k=(a_k, b_k)^T$. The momentum density has the following form
\begin{equation}
h(k)=h_0+h_x\sigma_x+h_y\sigma_y+h_z\sigma_z,
\end{equation}
where $h_0=\Delta_c$ and $\textbf{h}=\{h_x,h_y,h_z\}=
\{2J\cos(k_x),
2\delta\sin(k_x)-J_e\sin(\theta)\sin(k_x),J_e\cos(\theta)\}$ with
$J=(J_1+J_2)/2$, $\delta=(J_1-J_2)/2$ and $J_e=g^2_0/\Delta$. $\sigma_{x,y,z}$ are the
Pauli matrices spanned by $a_k$ and $b_k$. Interestingly, by
associating the mixing angle $\theta$ with the quasi-momentum $k_y$
in the second spatial direction, one can find that the above one dimensional circuit
QED lattice can be exactly mapped into a two dimensional Chern insulator Hamiltonian. As
plotted in Fig. 1(b), the $x$ direction quasimomentum $k_x$ and
the mixing angle $\theta$ can form a two dimensional
Brillouin zone $k_x\in[0,\pi]$ and $\theta\in[0,2\pi]$, which can
be rolled into a torus \cite{Kraus,Zhou,Lang,Zhu} for analyzing the
underlying topology in the artificial two dimensional lattice.

\begin{figure}[h]
\includegraphics[width=8cm,height=4cm]{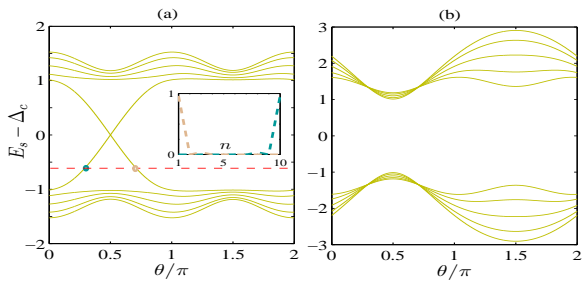}
\caption{Edge state spectrum for Photonic Chern insultaor with (a) Chern number $C=1$ when
$\delta=0$  and (b) Chern number $C=0$ when $\delta=0.6J_e$. For Chern insulator, there are two edge states at the in-gap energy
denoted by the red dashed line. Inset: the density distribution of
the two edge states. The other parameter
are chosen as $J=J_e$ and the lattice size $L=10$.  }
\end{figure}

 \emph{Photonic Chern insulator.} The topological properties of the model introduced above are captured by the Chern number of
 the Bloch band and the edge state spectrum. By mapping the two-dimensional torus to a spherical surface, the Chern number of
 the occupied ground band can be expressed as $C=\frac{1}{4\pi}\int\int
dk_xd\theta(\partial_{k_x}\hat{\textbf{h}}\times\partial_{\theta}\hat{\textbf{h}})\cdot\hat{\textbf{h}}$,
where the unit vector $\hat{\textbf{h}}=(h_x,h_y,h_z)/|h|$ with
$|h|=\sqrt{ h_x^2+h_y^2+h_z^2}$. Through substituting
$h_{x,y,z}$ into above formula and one can get the Chern number of
the ground band as
\begin{equation}
\label{cn}
    C=
   \begin{cases}
   1 &\mbox{if $-J_e<2\delta<J_e$}\\
   0 &\mbox{otherwise}
   \end{cases}
  \end{equation}
One can change the hopping difference $\delta$ to engineer the
photonic topological phase transition.  It is also worth pointing
out that, when the coupling strength $g_2=-g_0\cos(\theta/2)$, the
ground state can be prepared as a Chern insulator with $C=-1$. According to the bulk-edge correspondence, the appearance of edge
state is a hallmark of topological insulator. In Fig.
2(a), we have plotted the edge state spectrum for the topological
insulator. There is one pair of edge state at the in-gap energy.
The density distribution for the left and right edge states have
been plotted in the inset figure of Fig. 2(a). One finds that
there is one edge state localized at each edge.  For the
topological trivial insulator shown in Fig. 2(b), there is no edge
state in the gap.

\emph{Scattering formulation of topological invariant.} Based on
Laughlin's \cite{Laughlin} pumping argument, recent scattering
theory of  topological insulators shows that topological invariant
can be described by the reflection matrices at the Fermi level
\cite{scatp}. The basic experimental setup is achieved by rolling
a two-dimensional topological system into a cylinder and threading
it with a magnetic flux. For our one-dimensional photonic
simulator, if we regard the left and right edges of the photonic
lattice as the two ends of the cylinder, the periodic parameter
$\theta$ as the external magnetic flux and the
in-gap energy as the fermi level, our system can be naturally used to simulate the
experimental setup in Laughlin's pumping argument and to test
scattering theory of topological insulators. When the frequency of
the incident photon towards one edge is tuned into the in-gap
energy and the external periodic parameter $\theta$ is tuned over
one period, the pumping particle number per cycle can be expressed
as
\begin{equation}
\label{reflec} Q=\frac{1}{2\pi
i}\int^{2\pi}_0d\theta\,\frac{d}{d\theta}\,\text{log}\,r(\theta),
\end{equation}
where $r(\theta)$ is the reflection coefficient of the incident
photon from one edge. In this way, based on scattering theory of topological insulators \cite{scatp},
the topological invariant can be characterized by the winding number of the reflection
coefficient phase \cite{Hafezi20151}.

To further demonstrate this point, we use Green function to
analytically derive the reflection  coefficient from the left edge
of the above one-dimensional lattice (see supplemental materials),
giving
\begin{equation}
r(\theta)=-\frac{m_1+im_2}{m_1-im_2},
\end{equation}
where $m_1=4\delta
J+(E_p+J_e\cos(\theta))(\Delta_c+J_e\cos(\theta))-2JJ_e\sin(\theta)-
\sqrt{[E^2_p-4J^2-J^2_e\cos^2(\theta)][E^2_p-4\delta^2-J^2_e}$
$\overline{+4J_e\delta\sin(\theta)]}$,
$m_2=(E_p+J_e\cos(\theta))\sqrt{J^2-(E_p+\Delta_c)^2}$, $E_p$ is
the in-gap energy. By substituting the above equation into Eq.
(\ref{reflec}), we get
\begin{eqnarray}
Q&=&\frac{1}{2\pi}\int^{2\pi}_0d(\text{arctan}\,\frac{m^2_2-m^2_1}{2m_1m_2})\\ \nonumber
&=&\frac{1}{2}\,[\text{sgn}(2\delta+J_e)-\text{sgn}(2\delta-J_e)]\\
&=&C \nonumber.
\end{eqnarray}
One finds that the winding number of the phase of the reflection
coefficients is exactly equal to the topological  invariant of
this system. In the following section, we will show that the
information regarding the photonic reflection coefficient can be
probed spectroscopically using cavity input-output process. The
photonic Chern insulator is then detected by counting the winding
number of reflection coefficient phase.

\emph{Probing edge states and reflection coefficients.} In
contrast to fermi system,  one can directly probe the edge state
and its scattering feature in our photonic simulator. The reason
is that bosonic photons can occupy one particular eigenstate at
the same time. This could be done by externally driving the
resonators with the driven frequency tuned as the eigenenergy of
the lattice, then the corresponding eigenmode would be occupied
with some weights. In the rotating frame with respect to the
driving frequency, the driven Hamiltonian is
$H_d=\sum_{n}(\Omega_{na}a^{+}_n+\Omega_{nb}b^{+}_n)+h.c$, where
$\Omega_{na,nb}$ are the driven amplitudes in the $nth$ unit cell.
In the presence of dissipation, the expectation value of the
cavity field $a_j$ in steady state can be derived from the
solution of  the Lindblad master equation $\langle\dot{a_j}\rangle=-i\langle[a_j,H+H_d]\rangle+\kappa\sum_n\langle
L[a_n]a_j\rangle$, where the Lindblad term $L[a_n]a_j=a_n
a_ja^{+}_n-\{a^{+}_na_n,a_j\}/2$, $\kappa$ is the cavity decay
rate.  In the new bases $\vec{a}=(\langle a_1\rangle,\langle
b_1\rangle,...,\langle a_n\rangle,\langle b_n\rangle)^{Tr}$ and
$\vec{\Omega}=(\Omega_{1a},\Omega_{1b},...,\Omega_{na},\Omega_{nb})^{Tr}$
with $Tr$ representing the transposition of matrix, based on the
condition of the steady state solution
$\langle\dot{a_j}\rangle=0$, we can write the expectation value of
the cavity fields in the steady state as
\begin{equation}
\label{steady}
\vec{a}=-(\Delta_c+T-i\frac{\kappa}{2})^{-1}\vec{\Omega},
\end{equation}
where the elements of matrix $T$ are defined by
$T_{na,nb}=T_{nb,na}=J_1-J_e\sin(\theta)/2$,
$T_{na,(n-1)b}=T_{(n-1)b,na}=J_2+J_e\sin(\theta)/2$,
$T_{na(b),na(b)}=\pm J_e\cos(\theta)$.

\begin{figure}[h]
\includegraphics[width=9cm,height=8cm]{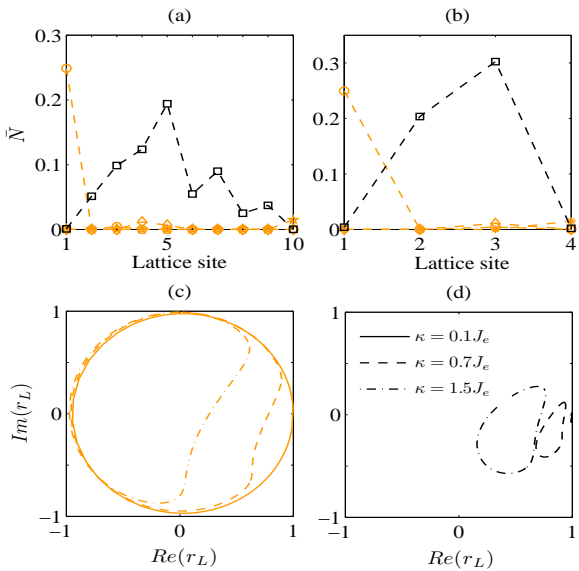}
\caption{ (a-b) The average photon number in the steady state when
the left edge state (circle) and one bulk state (square) are
driven to  be occupied by tuning  $\Delta_c=0.5J_e$ and $1.2J_e$,
respectively. For edge state case, the average photon numbers are
also plotted when the driven pulse is applied on the middle
(diamond) and rightmost (star) resonators. The reflection
coefficients from the left edge for topological (c) nontrivial and
(d) trivial insulators. The lattice size is (a,c,d) $L=10$ and (b)
$L=4$, the driving amplitude $\Omega_{1a}=0.1J_e$ and the cavity
decay rates $\kappa=0.1J_e$ (solid), $0.7J_e$ (dash) and $1.5J_e$
( dash-dot).}
\end{figure}

To probe the edge states, we need to occupy this edge states firstly.
As shown in Fig. 2(a), there is one pair of edge states at the
in-gap energy for photonic topological insulator. In
particular, we choose to excite the left edge state through
external driving the leftmost resonator (see Fig. 1(a)), with the
driven microwave pulses chosen as
$\vec{\Omega}=(\Omega_{1a},0,...,0,0)^{Tr}$ and driven frequency
$\omega_d$ tuned to the in-gap energy. The reason is that the left
edge state has maximal probability occupying the leftmost
resonator. In Fig. 3(a), we have plotted the average photon number
in the steady state for this case and find that the photons are
most localized in the left edge resonator. In contrast, if the
middle and rightmost resonators are driven with same laser, the
occupied probability of the left edge mode is very small, then
there will almost be no resonant eigenmode and all the photons
will finally decay into vacuum in the steady state. In contrast,
when the driven frequency is tuned as the bulk energy, the photons
are extensively populated in the lattice, which satisfies the
feature of Bloch bulk state. The results stay even if we choose
two unit cells (see Fig. 3(b)). Therefore, the photonic edge state
can be directly observed by measuring the corresponding average photon
number in the steady state.

The detection of photonic reflection coefficient is naturally related to
cavity input-output process \cite{QObook}.
Using input-output formalism, the reflected output
photons $a^{out}_1$ from the left edge
resonator is related to the input photon through
$a^{out}_1=a^{in}_1+\sqrt{\kappa}a_1$, where the input field
$a^{in}_1$ is related to the external driving by
$\sqrt{\kappa}a^{in}_1=i\Omega_{1a}$ \cite{rmpclerk}.
Using Eq. (\ref{steady}), the photonic reflection coefficient
from the left edge is obtained as
\begin{equation}
r_L(\theta)=\frac{\langle a^{out}_1\rangle}{\langle
a^{in}_1\rangle}=1+i\kappa[(\Delta_c+T-i\frac{\kappa}{2})^{-1}]_{11}.
\end{equation}
In Fig. 3(c) and (d), we plot the numerical results of reflection coefficients for photonic
topological nontrivial (Chern number $C=1$) and trivial insulator
(Chern number $C=0$). The results show that the winding number of the
reflection coefficient phase of $r_L$ is $1$ and $0$ respectively,
which yield the photonic topological invariants. This method also
applies for the right edge case and the conclusion is same.

Moreover, we also calculate the corresponding winding numbers even
when the lattice size is $L=4$ (two unit cells) and find that the
corresponding trajectories are same and all the results stay, which means that
topological states could be implemented with only a few-resonator lattice.
Such remarkable feature is quite attractive to circuit-QED experimenters. The current circuit QED experiment has already realized the
capacitive and qubit-assisted coupling between two resonators
\cite{tworesexp} (one unit cell). Compared with previous works,
our scheme is very promising and it is expected to be experimentally demonstrated.
In all cases, we
also take into account the influence of the cavity decay. The
results show that, if the cavity decay rate is not larger than the
energy gap $2J_e$, the in-gap energy will remain in the energy gap
and the winding number will remain the same, then our measurement
is very robust to fluctuations of the frequency of the input
photon.

\emph{Experimental discussion.} Before concluding, a detailed
estimate of  the experimental parameters involved is in order.
For circuit QED experiment \cite{tworesexp}, with a typical choice
of $\omega_d=5\Delta$, $g_0=0.1\Delta$, the qubit-assisted hopping
rate $J_e$ can approach the order of $10$ MHz. For the current
coupled transmission line resonator experiment \cite{circuitkoch},
the hopping rate $J_{1,2}$ can be tuned within the range $1-100$
MHz. One can easily check that the experimental parameters
required in our work are within the experimentally accessible regimes. For the
experimental detection of the reflection coefficient phase, we
assume an initial input driven lasers on the leftmost (rightmost)
resonator prepared in a coherent state $|\alpha\rangle$, then the
reflected output photon pulse will be in the coherent state
$||r_{L(R)}(\theta)|e^{i\phi}\alpha\rangle$, where $\phi$ is the
reflection coefficient phase. Based on homodyne detection by
interfering the output photons with a local oscillator, the phase
$\phi$ in the output coherent state $||r_{L(R)}|e^{i\phi}\alpha\rangle$
can be extracted. In this way, the winding number of the photonic
reflection coefficient phase is measured, yielding the photonic
topological invariant.

\emph{Summary.} In summary, we have introduced a conceptually simple method to realize a photonic Chern insulator in a one-dimensional circuit-QED
lattice. Based on Laughlin¡¯s pumping argument and input-output formalism, we have further demonstrated that the photonic edge states and topological invariant can be unambiguously measured even in a dissipative few-resonator network, which may take a significant step towards observing topological invariant with circuit-QED. Our method also provides a new route and simple means to study and probe high dimensional photonic topological states. By introducing
effective photon interaction, the exotic fractional photonic Chern insulator could be further studied in such framework.

\emph{Acknowledgements.} We thank M. Hafezi for his  comments. SLZ
was supported by the NSFC (Grant No. 11125417 and No.11474153),
the SKPBR of China (Grants No. 2011CB922104), the PCSIRT (Grant
No. IRT1243). JBY was supported by the NRF of Singapore (Grant No.
WBS: R-710-000-008-271). RF acknowledges financial support of
IP-SIQS. WN  and KLC acknowledge financial support of the National
Research Foundation, the Ministry of Education, Singapore.

\appendix

\begin{widetext}

\section{Tunable flux-qubit-resonator coupling}

\begin{figure}[h]
\includegraphics[height=6cm]{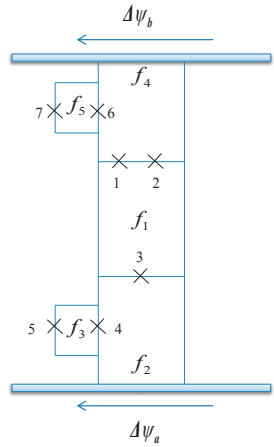}
\caption{Setup for tunable flux-qubit-resonator coupling.}
\end{figure}

 In each unit cell of our lattice system, the setup consists of two resonators coupled by one flux qubit. The flux qubit is made of three Josephson junctions. Using the method designed in \cite{Solano1,Solano2}, the coupling between resonators and flux qubit can be mediated by Josephson junctions. In the following, we will focus on the dominant energies of Josephson junctions $V(\phi_n)=-E_{Jn}\cos(\phi_n)$. Aussume the seven Josephson junction energies are $E_{J1}=E_{J2}=E_J$, $E_{J3}=\alpha E_J$, $E_{J4}=E_{J5}=\beta E_J$ and $E_{J6}=E_{J7}=\beta E_J$, each loop is applied by an external magnetic field with the flux $f_j=\Phi_j/\Phi_0$. Then the total potential energy of the whole system can be written as
 \begin{equation}
 U=-\sum^7_{n=1}E_{Jn}\cos(\phi_n)
 \end{equation}
 Based on the flux quantization condition
 \begin{eqnarray}
 \phi_1-\phi_2+\phi_3 &=& f_1   \\
 \phi_4-\phi_3+\Delta\psi_a &=& f_2  \\
 \phi_5-\phi_4 &=& f_3  \\
 \phi_2-\phi_1+\phi_6-\Delta\psi_b &=& f_4 \\
 \phi_7-\phi_6 &=& f_5
 \end{eqnarray}
 we can further rewrite the above potential as
 \begin{eqnarray}
 U &=& -E_J[\cos(\phi_1)+\cos(\phi_2)+\alpha\cos(f_1+\phi_2-\phi_1)   \\
 &+& 2\beta\cos(f_3)\cos(f_a+\phi_2-\phi_1+\Delta\psi_a)  \\
 &+& 2\gamma\cos(f_5)\cos(f_b-\phi_2+\phi_1-\Delta\psi_b)]
 \end{eqnarray}
 where $f_a=f_1+f_2+f_3/2$, $f_b=f_4+f_5/2$, the resonator fields are $\Delta\psi_a=\Delta\psi_{0}(a+a^{+})$ and
 $\Delta\psi_b=\Delta\psi_{0}(b+b^{+})$. \\

 By considering the resonator part and omitting the second order qubit-resonator coupling \cite{Solano1,Solano2}, based on the two degree of freedom $(\phi_1,\phi_2)$, one can get the following qubit-resonator Hamiltonian
 \begin{equation}
 H=\hbar\omega_0\sigma_z+\hbar\omega_c(a^{+}a+b^{+}b)
 +\sum_{\mu=x,y,z}[g_{a}c_{a\mu}(a+a^{+})+g_{b}c_{b\mu}(b+b^{+})]\sigma_{\mu}
 \end{equation}
 where $g_{a}=2\beta E_J\cos(f_3)\Delta\psi_{0}$ and $g_{b}=2\beta E_J\cos(f_5)\Delta\psi_{0}$, $c_{a\mu}$ and $c_{b\mu}$ are determined by the parameters $(\alpha,\beta,f_1,f_2,f_4)$. Similar to the method in \cite{Solano2}, the next step is to numerically find the parameters to make sure $c_{ay,az}=c_{by,bz}=0$ and $c_{ax,bx}=1$, then one can get
 \begin{equation}
 H=\hbar\omega_0\sigma_z+\hbar\omega_c(a^{+}a+b^{+}b)
 +[g_{a}(a+a^{+})+g_{b}(b+b^{+})]\sigma_{x}
 \end{equation}
 Note that the above Hamiltonian becomes the Jaynes-Cummings model after making rotating wave approximations. To further get the coupling modulations, the systems needs to work at
 \begin{eqnarray}
 f_5 &=& \theta/2 \\
 f_3 &=& (\pi-\theta)/2  \\
 f_5+f_3 &=& \pi/2
 \end{eqnarray}
 then we get the required periodic modulation on qubit-resonator couplings as
 \begin{equation}
 g_1=g_{a}=g_0\sin(\theta/2), g_2=g_{b}=g_0\cos(\theta/2)
 \end{equation}
 where $g_0=2\beta E_J\Delta\psi_{0}$. Actually, it is not necessary to have tunable qubit-resonator coupling strengths exactly with cosine and sine function forms. Instead, for the same purpose, one only needs to separately tune $g_1$ and $g_2$. Then, by introducing a mixing angle $\theta=2\arctan(g_1/g_2)$, $g_2$ and $g_1$ can be written in $\theta$ space with cosine and sine forms as shown in the above equation. \\

\section{Reflection coefficient}

In section, we give details using Green function to derive the reflection coefficient Eq. (10) in the main text. For this purpose, we model the total system in the scattering process into three parts: left lead, device (our system) and right lead. In particular, the reflection coefficient from the left lead is calculated below. The total lattice Hamiltonian of the system is $H=H_{\text{L}}+H_{\text{LD}}+H_{\text{D}}+H_{\text{RD}}+H_{\text{R}}$, where
\begin{equation}
\begin{split}
&H_{\text{D}}=\sum_{n=1}^{L/2}[(J_{1}-\frac{J_{e}}{2}\sin{\theta})a_{n}^{\dag}b_{n}+(J_{2}+\frac{J_{e}}{2}\sin{\theta})b_{n}^{\dag}a_{n+1}+\text{H.c.}]+\sum_{n=1}^{L/2}[J_{e}\cos{\theta}(a_{n}^{\dag}a_{n}-b_{n}^{\dag}b_{n})+\Delta_{c}(a_{n}^{\dag}a_{n}+b_{n}^{\dag}b_{n})],\\
&H_{\text{L}}=-\frac{J}{2}\sum_{i=-1}^{-\infty}(c_{i+1}^{\dag}c_{i}+\text{H.c.}),H_{\text{R}}=-\frac{J}{2}\sum_{i=L+1}^{\infty}(c_{i+1}^{\dag}c_{i}+\text{H.c.}).\\
\end{split}
\end{equation}
We assume the lattice sites of device $L$ is even. The tunnelings between leads and device are given by
\begin{equation}
\begin{split}
H_{\text{LD}}=-\frac{J}{2}(a_{1}^{\dag}c_{0}+\text{H.c.}),H_{\text{RD}}=-\frac{J}{2}(c_{L+1}^{\dag}b_{L}+\text{H.c.}).\\
\end{split}
\end{equation}
In the basis $\{\cdot\cdot\cdot,c_{-1}^{\dag},c_{0}^{\dag},a_{1}^{\dag},b_{1}^{\dag},\cdot\cdot\cdot,a_{L/2}^{\dag},b_{L/2}^{\dag},c_{L+1}^{\dag},c_{L+2}^{\dag},\cdot\cdot\cdot\}$, we can formulate the Hamiltonian of the whole system as
\begin{equation}
\begin{split}
H=\left[\begin{array}{*{20}ccc}
{H_{\text{L}}} & {\tau_{\text{L}}} & {0}\\
{\tau_{\text{L}}^{\dag}} & {H_{\text{D}}} & {\tau_{\text{R}}}\\
{0} & {\tau_{\text{R}}^{\dag}} & {H_{\text{R}}}\\
\end{array}\right],\\
\end{split}
\end{equation}
where
\begin{equation}
\begin{split}
&H_{\text{L}}=\left[\begin{array}{*{20}cccc}
{\cdot\cdot\cdot} & {\cdot\cdot\cdot} & {\cdot\cdot\cdot} & {\cdot\cdot\cdot}\\
{\cdot\cdot\cdot} & {0} & {-J/2} & {0}\\
{\cdot\cdot\cdot} & {-J/2} & {0} & {-J/2}\\
{\cdot\cdot\cdot} & {0} & {-J/2} & {0}\\
\end{array}\right]_{\infty\times\infty},H_{\text{R}}=\left[\begin{array}{*{20}cccc}
{0} & {-J/2} & {0} & {\cdot\cdot\cdot}\\
{-J/2} & {0} & {-J/2} & {\cdot\cdot\cdot}\\
{0} & {-J/2} & {0} & {\cdot\cdot\cdot}\\
{\cdot\cdot\cdot} & {\cdot\cdot\cdot} & {\cdot\cdot\cdot} & {\cdot\cdot\cdot}\\
\end{array}\right]_{\infty\times\infty},\\
&\tau_{\text{L}}=\left[\begin{array}{*{20}cccc}
{\cdot\cdot\cdot} & {\cdot\cdot\cdot} & {\cdot\cdot\cdot} & {\cdot\cdot\cdot}\\
{0} & {0} & {\cdot\cdot\cdot} & {0}\\
{0} & {0} & {\cdot\cdot\cdot} & {0}\\
{-J/2} & {0} & {\cdot\cdot\cdot} & {0}\\
\end{array}\right]_{\infty\times L},\tau_{\text{R}}=\left[\begin{array}{*{20}cccc}
{0} & {0} & {0} & {\cdot\cdot\cdot}\\
{\cdot\cdot\cdot} & {\cdot\cdot\cdot} & {\cdot\cdot\cdot} & {\cdot\cdot\cdot}\\
{0} & {0} & {0} & {\cdot\cdot\cdot}\\
{-J/2} & {0} & {0} & {\cdot\cdot\cdot}\\
\end{array}\right]_{L \times\infty},\\
&H_{\text{D}}=\left[\begin{array}{*{20}ccccc}
{\Delta_{c}+J_{e}\cos{\theta}} & {J_{1}-\frac{J_{e}}{2}\sin{\theta}} & {0} & {0} & {\cdot\cdot\cdot}\\
{J_{1}-\frac{J_{e}}{2}\sin{\theta}} & {\Delta_{c}-J_{e}\cos{\theta}} & {J_{2}+\frac{J_{e}}{2}\sin{\theta}} & {0} & {\cdot\cdot\cdot}\\
{0} & {J_{2}+\frac{J_{e}}{2}\sin{\theta}} & {\Delta_{c}+J_{e}\cos{\theta}} & {J_{1}-\frac{J_{e}}{2}\sin{\theta}} & {\cdot\cdot\cdot}\\
{0} & {0} & {J_{1}-\frac{J_{e}}{2}\sin{\theta}} & {\Delta_{c}-J_{e}\cos{\theta}} & {\cdot\cdot\cdot}\\
{\cdot\cdot\cdot} & {\cdot\cdot\cdot} & {\cdot\cdot\cdot} & {\cdot\cdot\cdot} & {\cdot\cdot\cdot}\\
\end{array}\right]_{L\times L}.
\end{split}
\end{equation}
Then the Green function for the device is given by \cite{Datta2005},
\begin{equation}
\begin{split}
G_{\text{D}}=[E\mathbf{I}-H_{\text{D}}-\Sigma_{\text{L}}^{r}-\Sigma_{\text{R}}^{r}]^{-1},\\
\end{split}
\end{equation}
where the self-energies of the leads are $\Sigma_{\text{L}}^{r}=\tau_{\text{L}}^{\dag}g_{\text{L}}^{r}\tau_{\text{L}}$ and $\Sigma_{\text{R}}^{r}=\tau_{\text{R}}g_{\text{R}}^{r}\tau_{\text{R}}^{\dag}$, and the lead Green functions are
\begin{equation}
\label{leadGF}
\begin{split}
g_{\text{L}}^{r}&=[(E+i\eta)\mathbf{I}-H_{\text{L}}]^{-1},\\
g_{\text{R}}^{r}&=[(E+i\eta)\mathbf{I}-H_{\text{R}}]^{-1}.\\
\end{split}
\end{equation}
After some straightforward calculations, we find that the non-zero elements in the self-energies are $[\Sigma_{\text{L}}^{r}]_{11}=\frac{J^2}{4}[g_{\text{L}}^{r}]_{\infty,\infty}$ and $[\Sigma_{\text{R}}^{r}]_{LL}=\frac{J^2}{4}[g_{\text{R}}^{r}]_{11}$, otherwise is zero. Due to the symmetrical configuration of the whole system, we note that $[\Sigma_{\text{L}}^{r}]_{11}=[\Sigma_{\text{R}}^{r}]_{LL}$.
Furthermore, the dispersion relation of the semi-infinite lead is $E=-J\cos{k}$, the group velocity in the lead is $\nu_{\text{L}}=\nu_{\text{R}}=\frac{\partial E}{\partial k}=J\sin{k}$ and the self-energy of the lead is $[\Sigma_{\text{L}}^{r}]_{11}=[\Sigma_{\text{R}}^{r}]_{LL}=-\frac{J}{2}e^{ika}$ \cite{Datta2005}. Keeping $\nu_{L}>0$ and $\nu_{R}<0$ for the photon injecting from leads to device, we have (assume $J>0$ from now on)
\begin{equation}
\begin{split}
\nu_{\text{L}}&=\sqrt{J^2-E^2}, \Sigma_{\text{L}}=[\Sigma_{\text{L}}^{r}]_{11}=\frac{1}{2}(E- i\sqrt{J^2-E^2}),\\
\nu_{\text{R}}&=-\sqrt{J^2-E^2}, \Sigma_{\text{R}}=[\Sigma_{\text{R}}^{r}]_{LL}=\frac{1}{2}(E+ i\sqrt{J^2-E^2}).\\
\end{split}
\end{equation}
Therefore, the device Green function is
\begin{equation}
\begin{split}
G_{\text{D}}=\left[\begin{array}{*{20}ccccc}
{E-\Delta_{c}-J_{e}\cos{\theta}-\Sigma_{L}} & {-J_{1}+\frac{J_{e}}{2}\sin{\theta}} & {0} & {0} & {0}\\
{-J_{1}+\frac{J_{e}}{2}\sin{\theta}} & {E-\Delta_{c}+J_{e}\cos{\theta}} & {-J_{2}-\frac{J_{e}}{2}\sin{\theta}} & {0} & {0}\\
{0} & {-J_{2}-\frac{J_{e}}{2}\sin{\theta}} & {\cdot\cdot\cdot} & {\cdot\cdot\cdot} & {0}\\
{0} & {0} & {\cdot\cdot\cdot} & {E-\Delta_{c}-J_{e}\cos{\theta}} & {-J_{1}+\frac{J_{e}}{2}\sin{\theta}}\\
{0} & {0} & {0} & {-J_{1}+\frac{J_{e}}{2}\sin{\theta}} & {E-\Delta_{c}+J_{e}\cos{\theta}-\Sigma_{R}}\\
\end{array}\right]_{L\times L}^{-1}.\\
\end{split}
\end{equation}
Via the continued fraction method and taking into account the periodic pattern of the matrix elements in $G_{\text{D}}$, the closed form of $[G_{\text{D}}]_{11}$ can be obtained,
\begin{equation}
\begin{split}
[G_{\text{D}}]_{11}^{-1}+\Sigma_{L}=E-\Delta_{c}-J_{e}\cos{\theta}-\frac{(J_{1}-\frac{J_{e}}{2}\sin{\theta})^2}{E-\Delta_{c}+J_{e}\cos{\theta}-\frac{(J_{2}+\frac{J_{e}}{2}\sin{\theta})^2}{[G_{\text{D}}]_{11}^{-1}+\Sigma_{L}}}.\\
\end{split}
\end{equation}
Solving this algebra equation we have
\begin{equation}
\label{GD11}
\begin{split}
[G_{\text{D}}]_{11}=-\frac{2(E_p+P_{1})}{m_{1}-im_{2}},\\
\end{split}
\end{equation}
where
\begin{equation}
\begin{split}
&m_{1}=J_{1}^2-J_{2}^2+(E_p+P_{1})(\Delta_{c}+P_{1})-(J_{1}+J_{2})P_{2}-\sqrt{[E_p^2-(J_1+J_2)^2-P_{1}^2][E_p^2-(P_{2}-J_{1}+J_{2})^2-P_{1}^2]},\\
&m_{2}=(E_p+P_{1})\sqrt{J^2-(E_p+\Delta_{c})^2},\\
&P_{1}=J_{e}\cos{\theta}, P_{2}=J_{e}\sin{\theta},\\
\end{split}
\end{equation}
and $E_p=E-\Delta_{c}$ is the in-gap energy of our photonic system. Based on the Fisher-Lee relation \cite{Fisher-Lee}, $S_{nm}(E)=-\delta_{nm}+i\sqrt{\nu_{n}\nu_{m}}[G_{\text{D}}]_{nm}$, where the scattering matrix is $S=\left[\begin{array}{*{20}ccc}
{r_{\text{L}}} & {t_{\text{R}}}\\
{t_{\text{L}}} & {r_{\text{R}}}\\
\end{array}\right]$. The reflection coefficient from the left lead is thus
\begin{equation}
\label{rL}
\begin{split}
r_{L}(\theta)=-1+i\sqrt{J^2-E^2}[G_{\text{D}}]_{11}.\\
\end{split}
\end{equation}
 Therefore, through substituting Eq. (\ref{GD11}) into Eq. (\ref{rL}), we find that
\begin{equation}
\begin{split}
r_{L}(\theta)=-\frac{m_{1}+im_{2}}{m_{1}-im_{2}}.\\
\end{split}
\end{equation}

\end{widetext}

\end{document}